\documentclass[3p,twocolumn]{elsarticle}

\usepackage{graphicx,amsmath,amssymb}
\usepackage{url}

\newcommand{\bfr}{\mathbf{r}}
\newcommand{\bfep}{\mathbf{e}_\psi}
\newcommand{\cslpm}{c_{\rm sLPM}}
\newcommand{\casw}{c_{\rm ASW}}
\newcommand{\ncoll}{n_{\rm coll}}

\begin{document}

\begin{frontmatter}

\title{Event-by-Event Jet Quenching}

\author[tamu]{R.\ Rodriguez}

\author[tamu,rbrc]{R.\ J.\ Fries}
\ead{rjfries@comp.tamu.edu}

\author[utep]{E.\ Ramirez}

\address[tamu]{Cyclotron Institute and Physics Department, Texas A$\&$M
University, College Station, TX 77843}
\address[rbrc]{RIKEN/BNL Research Center, Brookhaven National Laboratory,
Upton, NY 11973}
\address[utep]{Physics Department, University of Texas 
El Paso, El Paso TX 79968}

\begin{abstract}
High momentum jets and hadrons can be used as probes for the quark gluon
plasma (QGP) formed in nuclear collisions at high energies. We investigate the
influence of fluctuations in the fireball on jet quenching observables
by comparing propagation of light quarks and gluons through averaged, smooth 
QGP fireballs with event-by-event jet quenching using realistic inhomogeneous 
fireballs. We find that the transverse momentum and impact parameter 
dependence of the nuclear modification factor $R_{AA}$ can be fit well
in an event-by-event quenching scenario within experimental errors.
However the transport coefficient $\hat q$ extracted from fits to the 
measured nuclear modification factor $R_{AA}$ in averaged fireballs 
underestimates the value from event-by-event calculations by up to 50\%. 
On the other hand, after adjusting $\hat q$ to fit $R_{AA}$ in the 
event-by-event analysis we find residual deviations in the azimuthal
asymmetry $v_2$ and in two-particle correlations, that provide a possible 
faint signature for a spatial tomography of the fireball. We discuss a
correlation function that is a measure for spatial inhomogeneities in a 
collision and can be constrained from data.
\end{abstract}

%\pacs{12.38.Mh;24.85.+p;25.75.-q}

\end{frontmatter}

\section{Introduction}

Collisions of nuclei at high energies at the Relativistic Heavy
Ion Collider (RHIC) and, soon, at the Large Hadron Collider (LHC)
create a fireball with local energy densities well above 1 GeV/fm$^3$. At
those densities, quarks and gluons form a deconfined quark gluon plasma 
(QGP) \cite{whitepaper:05}.
In some of the collisions, high momentum partons in the initial nuclear 
wave functions scatter off each other and propagate away from the 
collision axis. They form large momentum jets in the final state. 
These jets, and the hadrons fragmenting from them, can be used as hard 
probes of the fireball. The interactions of the scattered partons 
with quark gluon plasma lead to radiative energy loss and a significant 
suppression of the hadron yield at high transverse momentum 
$P_T$ \cite{Wang:1991xy,BDMPS:96,Zakharov:96,Wiedemann:2000tf,gyulassy,AMY:02,Wang:2003aw}. 
One of the key results from RHIC was the confirmation of this jet quenching
effect: Hadrons with $P_T \geq 5$ GeV/$c$ are suppressed by about 
a factor 5. In addition, an extinction of away-side jet 
correlations has been seen in a certain kinematic regime, further
emphasizing the large opacity of quark gluon plasma \cite{whitepaper:05}.

The study of quark gluon plasma has now moved into an era of quantitative
assessments of experimental results. A simple question that we should
be able to answer is that of an averaged value $\langle \hat q\rangle$
of the transport coefficient
%\begin{equation}
  $\hat q = \mu^2/\lambda$,
%\end{equation}
the average squared momentum transfer $\mu^2$ of a high momentum parton 
per mean-free path $\lambda$. The averaging $\langle\ldots\rangle$ here 
refers to the many possible paths of a parton (thus sampling 
different local $\hat q$ along the trajectory in a cooling and expanding 
fireball), an average over parton species (until we have means to reliably 
distinguish gluon and quark jets), and the average over many event 
geometries for a given centrality bin (hard processes at RHIC are rare
and experimental results are event averaged). 

Comparative studies using the mainstream energy loss models lead to
a somewhat unsettled picture. Bass et al.\ \cite{Bass:2008rv} have reported 
a wide range of possible values for $\hat q_0$, the local initial value of
$\hat q$ at the center of a central collision, ranging from 2.3 GeV$^2$/fm
to 18.5 GeV$^2$/fm depending on the energy loss model, and on how 
the local $\hat q (\mathbf{r},\tau)$ is modeled as a function of the local 
energy density or temperature at position $\mathbf{r}$ and time $\tau$.
In addition, we have been cautioned by  results that show that 
$\hat q$ extracted from single and di-hadron nuclear suppression factors 
are not necessarily compatible, and that $\hat q$ is sensitive to 
assumptions on pre-equilibrium quenching and the initial parton spectrum
\cite{Armesto:2009zi} as well as radiative corrections to the hard 
process \cite{Zhang:2007ja}.

Clearly we have to discuss and constrain uncertainties in our modeling
of jet quenching very carefully in order to arrive at reliable quantitative
estimates. In this work we investigate the influence of inhomogeneities 
and fluctuations in the fireball on jet quenching observables. As mentioned
above, experimental data are averages of observables over many events,
where in each event jets are created and propagate through the underlying
fireball. Most calculations found in the literature on the other hand
turn this process around and propagate jets through an 
idealized fireball which can be understood as an average over
realistic fireballs. These smooth fireballs come in various
degrees of sophistication, from a simple overlap of nuclear thickness functions
in the transverse plane to the use of detailed maps from hydrodynamics
calculations that take into account the proper expansion and cooling.
However, even the latter are often based on averaged and hence idealized
initial conditions. It has been realized before that event-by-event
computations are crucial to understand some low-$P_T$ observables, 
e.g.\ hydrodynamic elliptic flow \cite{Miller:2003kd}. It is important 
to take into account that the overlap of two nuclei (i) is irregularly 
shaped, (ii) is generally not aligned with the naive geometrical reaction 
plane, and (iii) exhibits local fluctuations with hot spots and cooler 
regions. This leads to appreciable differences compared to computations 
using averaged, idealized fireballs.

Here we investigate whether an event-by-event computation of jet quenching
differs from one using an averaged event. If the answer is yes, 
an interesting question arises: is it possible to study 
some features of the spatial structure of fireballs with hard probes, 
despite the averaging over events? In other words, is 
a true tomography feasible?

\section{Effects of Inhomogeneities on Quenching}

Let us discuss some general expectations when we go from parton 
propagation through an averaged fireball to an average over propagation in
many fireballs. First consider the limit of extreme quenching,
$\hat q R^2 \gg p$ where $R$ is the typical size of the fireball and $p$
the momentum of the final state parton. All observed particles then come 
from the surface of the fireball. It is clear that we should expect more 
such particles from an inhomogeneous fireball compared to a smooth 
fireball with equal total energy, if $\hat q$ is a fixed function of the
fireball density. This is due
to the larger effective surface area of an inhomogeneous fireball, 
see e.g.\ Fig.\ \ref{fig:events}. Hence the 
single and double particle nuclear modification factors, 
\begin{align}
  R_{AA}(P_T) &= \frac{dN^{AA}/dP_T}{\langle N_{\rm coll}\rangle dN^{pp}/dP_T}\\
  J_{AA}(P_{T1},P_{T2}) &= \frac{dN^{AA}/dP_{T1}dP_{T2}}{\langle N_{\rm coll}\rangle 
  dN^{pp}/dP_{T1} dP_{T2}}
\end{align}
should increase for partons, and for hadrons fragmenting from them.
We also expect the azimuthal anisotropy
\begin{equation}
  v_2 (P_T) = \frac{\int d\psi \cos(2\psi) dN^{AA}/dP_T d\psi}{
  \int d\psi dN^{AA}/dP_T d\psi} \, ,
\end{equation}
i.e.\ the difference of parton emission out of the reaction plane and into 
the reaction plane, to decrease since 
the relative increase in surface should be larger on the out-of-plane side.
Please note that the number of collisions $N_{\rm coll}$ in the denominator
is an averaged number estimated for the corresponding centrality bin.
We will follow this procedure and will not divide by the number
of collisions on an event-by-event basis. For completeness let us
also give the definition for the nuclear modification factor of the 
two-particle correlation per trigger 
$I_{AA}(P_{T1},P_{T2}) = J_{AA}(P_{T1},P_{T2})/R_{AA}(P_{T1})$ which we will
use later.
%However, there could in principle be a systematic bias 
%from event-by-event fluctuations of $N_{\rm coll}$.

Let us consider a more quantitative example. Imagine energy
loss $\Delta E = C h_\beta$ along a parton trajectory determined by an 
expression of the general type
\begin{equation}
  \label{eq:h}
  h_{\beta}(\bfr,\psi) = \int d\tau \tau^\beta \rho(\bfr+\tau \bfep)
\end{equation}
where $\bfr$ and $\psi$ are the point of creation and the emission angle
of the parton, $\bfep$ the unit vector along the trajectory, and $\tau$
the time elapsed since creation of the parton. $\beta$ encodes the 
path-length dependence with linear or quadratic dependence 
corresponding to $\beta=0$ or $\beta=1$ resp., and $C$ is a coefficient.
$\rho(\bfr)$ encodes a local property of the fireball, akin to a density,
which we do not specify further at this point. Let $n(\bfr)$ 
be the probability for a parton to emerge from point $\bfr$.
The relevant quantity to study is the energy loss weighted with the emission 
probability, $n(\bfr)h_\beta(\bfr,\psi)$. This quantity characterizes the 
suppression of single particle spectra for not too large quenching 
($p \gg \Delta E$) as we can infer from the following exercise for a 
power-law parton spectrum $dN^{\rm init}/d^2 \mathbf r d\psi dp \sim  n(\bfr) p^{-\alpha}$.
Expanding the expression $(p+C h_\beta)^{-\alpha}$ in the final parton spectrum
for small energy loss we can express
%\begin{multline}
%  \frac{dN^{\rm final}}{d^2 \bfr dp d\psi} = 
%  (p+ Ch_{\beta}(\bfr,\psi) )^{-\alpha} \\ = 
%  p^{-\alpha} - C \alpha p^{-\alpha-1} h_{\beta} (\bfr,\psi) + \ldots
%\end{multline}
the final spectrum as
\begin{multline}
  \frac{dN^{\rm final}}{dp d\psi}
  =  \frac{dN^{\rm init}}{dp d\psi} \\  - \alpha p^{-\alpha-1}
     \int d^2 \bfr n(\bfr) h_{\beta}(\bfr,\psi) + \ldots   \, .
\end{multline}
The energy loss model here is deterministic, but we do not expect major 
modifications of the following arguments if $h_{\beta}$ only gives an 
average value of a statistical process of scattering and gluon emission.

Now we consider the pair of densities $n_i(\bfr)$, $\rho_i(\bfr)$ 
event-by-event by writing them as a sum of ensemble expectation
values $\bar n(\bfr)$, $\bar \rho(\bfr)$ and fluctuations $\delta n(\bfr)$, 
$\delta \rho(\bfr)$, resp.
The ensemble average of the single particle suppression is given by
\begin{multline}
  \langle n(\bfr)h_{\beta}(\bfr,\psi)\rangle = \bar n(\bfr) \int d\tau \tau^\beta 
  \bar \rho(\bfr,\psi) \\
%  + \langle \delta n (\bfr)\rangle  \int d\tau \tau^\beta 
%  \bar \rho(\bfr +\tau \bfep) 
%  + \bar n (\bfr) \int d\tau \tau^\beta \langle \delta \rho(\bfr +\tau \bfep)
%  \rangle
  + \int d\tau \tau^\beta \langle \delta n(\bfr)
  \delta \rho (\bfr +\tau \bfep)\rangle \, .
\end{multline}
The first term is the result from propagating through the averaged fireball,
and we have omitted terms linear in fluctuations due to 
$\langle \delta n\rangle = 0$, $\langle \delta \rho \rangle = 0$. 
The last term contains the correction due to fluctuations
\begin{equation}
 \label{eq:dnh}
 \delta(nh_{\beta})(\bfr,\psi) = \int d\tau \tau^\beta R(\bfr,\bfr+\tau\bfep) \, .
\end{equation}
We have introduced the correlation function 
\begin{equation}
  \langle \delta n(\bfr_1)\delta \rho(\bfr_2) \rangle = R(\bfr_1,\bfr_2)
\end{equation}
between fluctuations of the position of hard collisions and the density of
the bulk fireball. Eq.\ (\ref{eq:dnh}) is a rather general statement one can 
make about event-by-event fluctuations without imposing too many restrictions
on the energy loss mechanism. Our result indicates that the leading
deviation due to fluctuations is given by correlations between the 
emission point of the jet and the fireball along its trajectory.

\begin{figure}[tb]
\centerline{\includegraphics[width=\columnwidth]{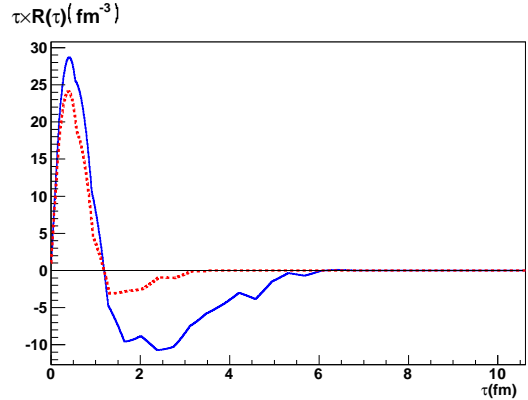}}
%\centerline{\includegraphics[width=\columnwidth]{3D_density_smooth_scaled.eps}}
\caption{The correlation function 
$\tau R(\tau)=\tau R(\bfr,\bfr + \tau \bfep)$ as a function of 
$\tau$ along two cuts extending radially from the point 
$\bfr=4\>\mathrm{fm}\> \mathbf{e}_y$, 4 fm away from the reaction plane, 
in $x$ ($\psi=0$, solid line) and $y$ direction ($\psi = \pi/2$, dashed line)
resp.; calculated 
from 10,000 GLISSANDO Au+Au events with an average impact parameter 
$b=3.2$ fm.}
\label{fig:rr}
\end{figure}

What constraints can be put on $R$? We expect fluctuations to 
be granular with a certain length scale $\sigma$ (e.g.\ the nucleon diameter 
if $n$ is related to the density of nucleon-nucleon collisions and
$\rho$ to the density of participant nucleons). Then $R$ must be positive for
$\left|\bfr_2-\bfr_1\right| \lesssim \sigma$ because on average the density
of hard processes and the density of the soft, ``underlying'' event
should be positively correlated. On the other hand, $R$ will turn negative 
on distance scales larger than $\sigma$ because the total amount of matter 
in the transverse plane is conserved on average. In other words, a hot 
spot of the fireball has to be compensated, on average, by less material 
around that spot. This is also the reason for the argument of a larger 
effective surface that we raised for the case of surface-dominated emission: 
Clumping of density somewhere along the boundary of the fireball introduces 
``holes'' elsewhere.
In principle $\rho$ also carries an explicit dependence on the time $\tau$
elapsed --- suppressed in the notation --- since the fireball 
is evolving dynamically. We expect that relaxation phenomena or 
hydrodynamic evolution wash out the correlation function $R$, 
although this might take several fm/$c$ by which time a large fraction 
of the observable jet strength has left the fireball.
We check our qualitative expectations with an example shown in Fig.\
\ref{fig:rr}. We provide two cuts through the correlation function of the 
densities of binary collisions, $\rho=n\equiv\ncoll$, calculated from the
initial state simulation GLISSANDO \cite{Broniowski:2007nz}. We discuss 
more details about GLISSANDO in the next section. We clearly see positive 
correlations with a radius $\sigma \approx 1$ fm as expected from fluctuations 
based on collisions of nucleons. The anti-correlation region extends all 
the way to the point where the nuclear overlap zone ends.

For the following it is useful to look at a simplistic parametrization for 
the correlation 
function $R$ which should be qualitatively true for a wide class of models.
Let us assume a non-spherical fireball with short and long axes $X$ and $Y$,
resp. In a fluctuating fireball these are only expectation values, of course. 
Let us further assume 
\begin{equation}
  R(\bfr_1,\bfr_2) \\ = \lambda \Theta(\sigma -\Delta r) 
    - \mu \Theta(\Delta r-\sigma)
\end{equation}
where $\Delta r = |\bfr_2-\bfr_1|$ and we neglect the dependence of $R$ 
on the center coordinate $\bfr_1+\bfr_2$ which should be a satisfying 
approximation for fireballs which are \emph{on average} uniform 
(i.e. $\bar \rho$ and $\bar n$ are smooth) and large, (i.e.\ $X$, $Y \gg 
\sigma$). $\lambda$ and $\mu$ are positive numbers that characterize 
the correlation strength on distance scales $\sigma$ and the 
anti-correlation strength on larger distances, resp. We note that $R$ 
should go to zero if the relative distance becomes too large which 
is not duly captured in the ansatz above. However this should not change 
the following interesting result on elliptic flow. First we note that
\begin{equation}
 \delta(nh)(\bfr,\psi) \approx \lambda \sigma^{\beta+1} - \mu \left( 
 l^{\beta+1}(\bfr,\psi) - \sigma^{\beta+1} \right)
\end{equation}
where $l$ is the length of the parton trajectory in the fireball.
We clearly see that the sign of the correction is determined by
the competition between an increased suppression coming from more jets 
being emitted in regions with a denser fireball, and the decreased suppression
around those regions.

While it is hard to predict the sign of $\delta(nh)$ even
after integration over emission points $\bfr$ without any further concrete
assumptions we can make the following observation. Let us use
the difference of energy loss in- and out-of-plane as a proxy for $v_2$.
To be more precise, $v_2$ should be a monotonously rising function of
\begin{equation}
  -\int d^2 \bfr \, n(\bfr) \left( h(\bfr,0) - h(\bfr,\pi/2) \right) \, .
\end{equation}
Under the assumptions made here the correction to this asymmetry
due to fluctuations is
\begin{multline}
 -\int d^2 \bfr \left( \delta(nh)(\bfr,0) - \delta(nh)(\bfr,\pi/2) \right) \, .
 \\ \approx \mu \int d^2 \bfr  \left( l^{\beta+1}(\bfr,0) -
 l^{\beta+1}(\bfr,\pi/2) \right)
 \\ \sim X^{\beta+2}Y - Y^{\beta+2}X < 0
\end{multline}
for reasonable $\beta$ since $X<Y$.
Hence, the azimuthal anisotropy $v_2$ tends to be diminished in event-by-event
calculations for a broad variety of energy loss models. Basically 
there is more room for the anti-correlation in $R$ to decrease energy 
loss along the longer side of the fireball than along the narrow side. 
This is compatible with the argument we made in the case of extremely 
strong quenching and surface dominated emission, and it can also be seen
in Fig.\ \ref{fig:rr}.

\section{Numerical Study}

We want to back up some of the analytic arguments from the last section 
through a numerical study.
The distribution of hard collisions is usually taken to be the density
of binary nucleon-nucleon collisions, $n(\bfr) = \ncoll(\bfr)$.
$\hat q(\bfr) $ is often assumed to be a function of the local energy 
density $\epsilon(\bfr)$ in the transverse plane. Around midrapidity the 
initial energy density is usually modeled as a superposition of the density 
of collisions and the density of participant nucleons 
$\epsilon(\mathbf{r}) = \alpha n_{\rm part}(\mathbf{r}) +\gamma
\ncoll(\mathbf{r})$.
%In the simplest (event averaged) case the density of collisions and 
%participants would be given by the product and the sum, resp.,  of the 
%average nuclear thickness functions $T_A(\mathbf{r}+\mathbf{b}/2)$ and 
%$T_B(\mathbf{r}-\mathbf{b}/2)$ of the two nuclei $A$ and $B$ where 
%$\mathbf{b}$ is the impact vector defining the reaction plane together 
%with the beam axis. 
%Using averaged thickness functions can only procedure expectation values 
%for the fireball averaged over many collisions.
%The time evolution of $\epsilon(\bfr)$ can be modeled by hydrodynamics.

\begin{figure}[tb]
%\centerline{
%\includegraphics[width=\columnwidth,%height=3.6in,angle=-90
%]{2_D_density_nonsmooth_3.2.eps}}
%\hspace{-0.2cm}
%\centerline{\includegraphics[width=\columnwidth]{2_D_density_smooth_3.2.eps}}
\centerline{\includegraphics[width=\columnwidth]{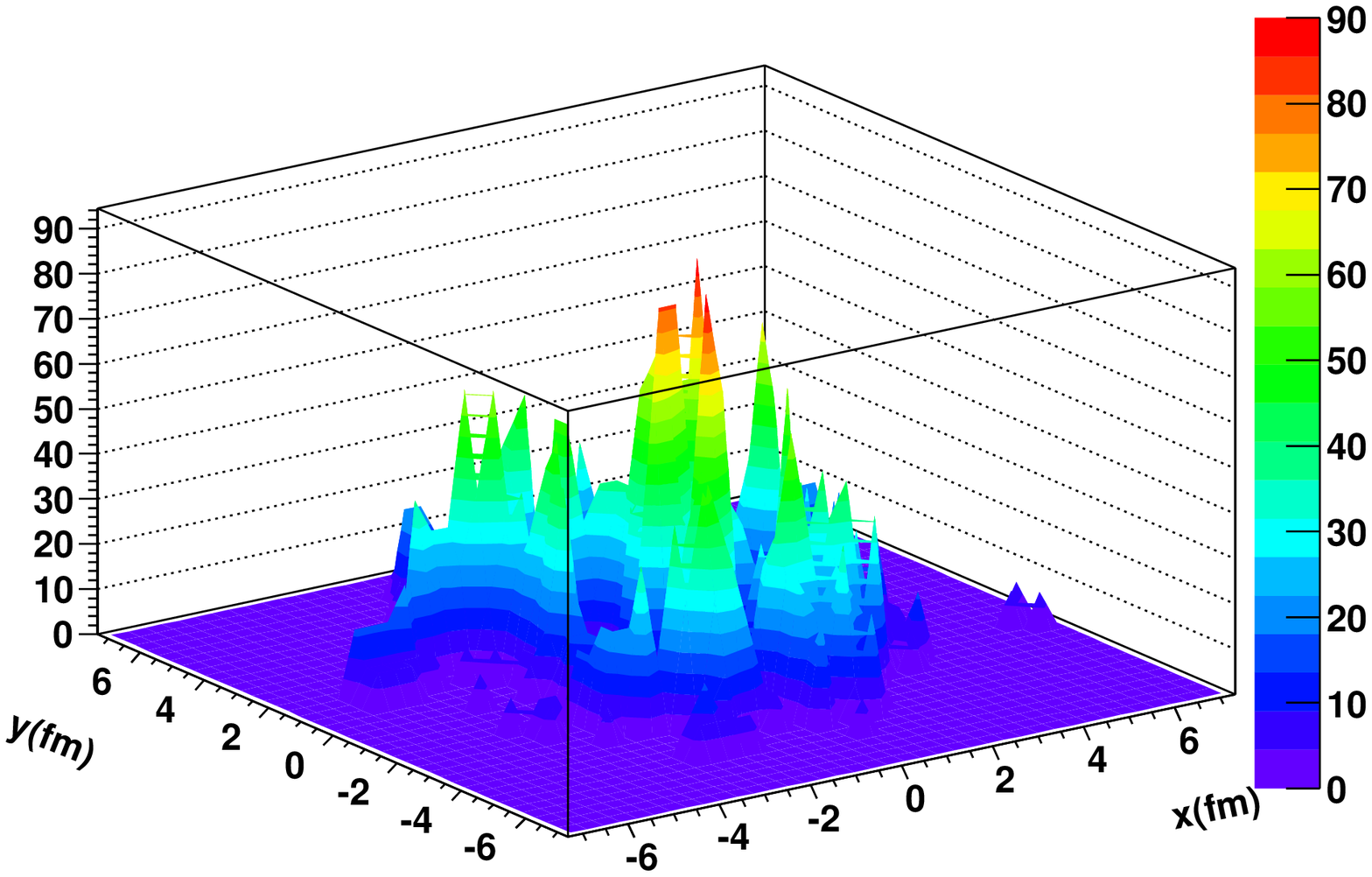}}
\centerline{\includegraphics[width=\columnwidth]{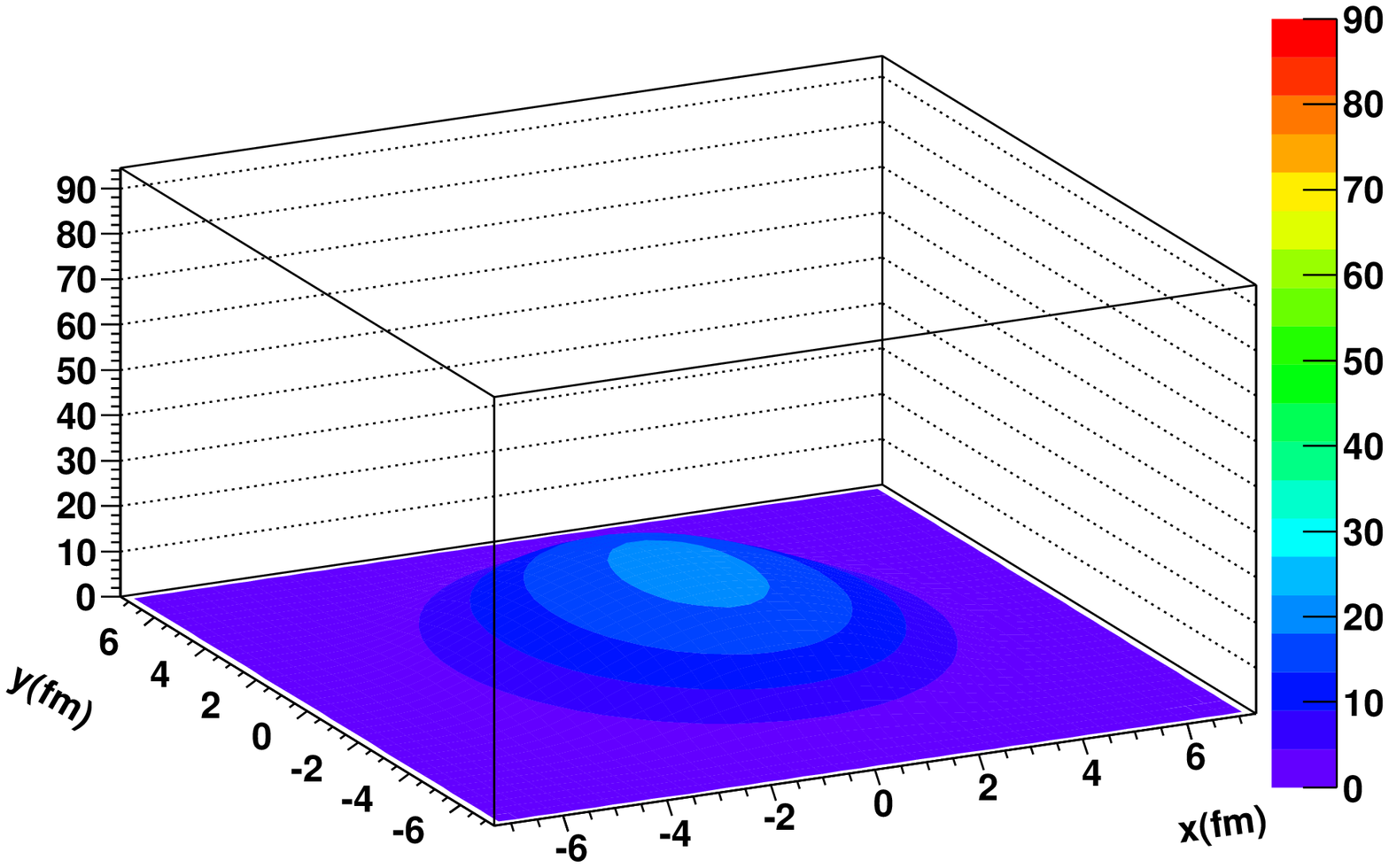}}
\caption{Top panel: density of binary collisions in the transverse plane for 
a typical GLISSANDO event for a centrality bin around impact parameter 
$b=3.2$ fm. Lower panel: the same averaged over 500,000 such events. 
The total number of collisions for the particular individual event here 
is about 15\% larger than the average number of collisions for this bin.}
\label{fig:events}
\end{figure}

Here, we produce an ensemble of realistic initial distributions through the
Glauber-based event generator  GLISSANDO \cite{Broniowski:2007nz}.
We take $n= \ncoll$ as above and for simplicity identify the initial 
density of the fireball as $\rho(\bfr)\sim \ncoll(\bfr)$ as well,
as in some well-known energy loss model calculations \cite{Dainese:2004te}. 
We do not implement a time evolution, 
since we only look at deviations of observables from their counterparts 
in smooth, averaged collisions. In other words we are only sensitive to 
the time evolution of $R(\bfr_1,\bfr_2)$.
However we can argue that the longitudinal expansion of the fireball
will not change the transverse correlation function $R$ except for 
an overall scaling factor, and transverse expansion is building up from
zero at early times, being not overly relevant for most measured jets.
Smooth fireballs are created by averaging over
500,000 GLISSANDO events in the corresponding centrality bin.
Fig.\ \ref{fig:events} compares a typical single
event around $b=3.2$ fm with the averaged event of the same
centrality. The highly 
fragmented nature of this fireball is evident. We use
GLISSANDO with the default values provided \cite{Broniowski:2007nz}.
All our runs have the following choices made: binary collisions, 
no superimposed weights, and variable-axes quantities. To make contact
between a range of impact parameters $b$ in GLISSANDO and experimental
centrality bins we use the tables in \cite{Reygers:2004}.

We use the software package PPM to calculate jet quenching results. 
PPM is a modular code developed by us to calculate hard probes observables. 
Here we run it in a mode
that propagates samples of hard partons on eikonal trajectories through the
background fireball with different leading particle energy 
loss models selected. We let PPM read in GLISSANDO output for both 
the distribution of hard processes and as a map for the
fireball for a given event. The initial momentum distribution of quark and 
gluon jets used follows a leading order pQCD calculation 
\cite{Owens:1986mp,fries1}. As we check against pion data PPM uses the option 
for KKP fragmentation \cite{akk} which gives reasonable results for pions.
Finally PPM computes $R_{AA}$, $I_{AA}$, and $v_2$.
For leading particle energy loss we espouse two options in PPM: 
(i) a simple, deterministic,
LPM-inspired model (sLPM) in which $\Delta E = \cslpm h_1$ where
$h_1$ is given by Eq.\ (\ref{eq:h}). The parameter $\cslpm$ measures the 
relative quenching strength $\cslpm = \hat q(\bfr)/\ncoll(\bfr)$.
(ii) the energy loss model known as the Armesto-Salgado-Wiedemann
(ASW) formalism, which is non-deterministic. Instead, it assigns a 
probability density for energy loss which is given as
\cite{Salgado:2003gb}
\begin{equation}
P(\Delta E;R,\omega_c)=p_{0}\delta(\Delta E)+p(\Delta E;R,w_c) 
\end{equation}
where $p_0$ is the probability to have no medium-induced gluon radiation and 
the continuous weight $p(\Delta E )$ is the probability to radiate an 
energy $\Delta E$ if at least one gluon is radiated. 
%The values of $p_0$ and 
%$p(\Delta E )$ are known as the quenching weights. 
In order to find these two quantities for each trajectory in our 
fireball we define $\rho(\bfr) = \casw \ncoll$, and PPM computes the integrals 
$h_1(\bfr, \psi)$ and $h_2(\bfr, \psi)$. The probability distributions
are evaluated in the multiple soft scattering approximation 
(the ASW-BDMPS formalism) \cite{Salgado:2003gb}, by using the relations 
introduced in \cite{Dainese:2004te} :
\begin{equation}
  \omega_c= h_1  \mbox{  and  } R = 2 h^2_{1}/h_0  \, .
\end{equation}
For the Monte Carlo sampling of the distribution we choose the 
non-reweighting algorithm explained in Ref.\ \cite{Dainese:2004te}.
As in scenario (i) the parameter $\casw$ gives the quenching strength
per density.

 %Observe that $I_1$ corresponds to $\beta=1$ in (\ref{eq:h}) 

\begin{figure}[tb]
\centerline{\includegraphics[width=\columnwidth]{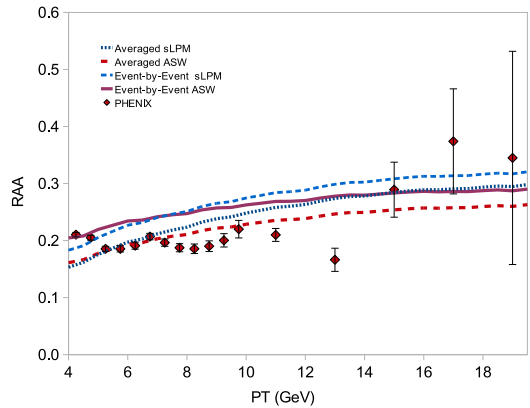}}
%\centerline{\includegraphics[width=\columnwidth]{3D_density_smooth_scaled.eps}}
\caption{$R_{AA}$ of neutral pions for $b$ around 3.2 fm computed for 
sLPM and ASW energy loss compared with PHENIX data \cite{Adare:2008qa}. 
Both average fireball and event-by-event calculations are shown, using the
same values of $\cslpm$ and $\casw$.}
\label{fig:raa1}
\end{figure}

We fit the energy loss parameters $\cslpm$ and $\casw$ by comparing 
PHENIX data on neutral pion suppression $R_{AA}$ at top RHIC energy for 
three different centralities: $0-10\%$, $20-30\%$ and $50-60\%$ 
\cite{Adare:2008qa} to PPM calculations using averaged fireballs 
for three corresponding impact parameter bins. The extracted values are
$\cslpm=0.055$ GeV and $\casw=1.6$ GeV. We note that ASW requires a much
larger relative quenching, but we do not want to focus on a comparison
of different energy loss models here. We simply use two models to estimate 
the uncertainties associated with our incomplete attempts to quanitify energy
loss, and we only focus on relative changes between the smooth and the 
event-by-event case.

Next we run PPM over samples of individual events and then take the average 
of our observables, keeping $\cslpm$ and $\casw$ constant.
For all centralities and all values of $P_T$ we observe an \emph{increase} in 
$R_{AA}$, i.e.\ a consistently lower energy loss event-by-event
compared to results using an averaged fireball.
Fig.\ \ref{fig:raa1} compares both scenarios and PHENIX data for central 
collisions (around $b=3.2$ fm) using both the sLPM and ASW energy loss.
The deviations grow going from central to peripheral collisions.

\begin{figure}[tb]
\centerline{\includegraphics[width=\columnwidth]{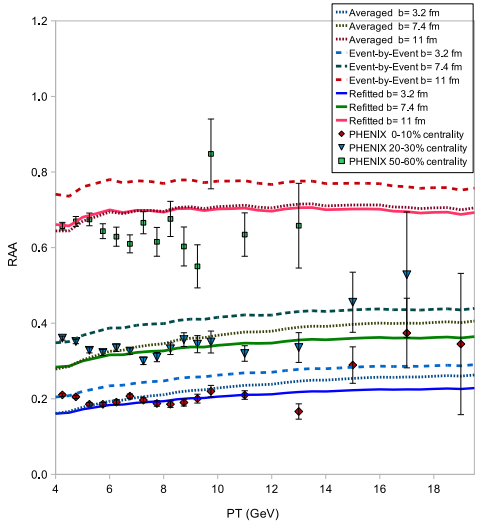}}
\caption{$R_{AA}$ of neutral pions for three centrality bins computed
in the ASW energy loss model compared to PHENIX data \cite{Adare:2008qa}. For each 
bin we show the result for the averaged fireball (dotted lines) for
$\casw=1.6$ GeV, for the event-by-event computation (dashed lines) for the
same quenching strength, and the event-by-event calculation for the
refitted value of $\casw=2.8$ GeV (solid lines).
}
\label{fig:raa2}
\end{figure}

Now we check whether the decreased suppression can be absorbed in a 
redefinition of the quenching strength. Indeed, at not too large 
transverse momentum
we find that calculations of $R_{AA}$ using event-by-event quenching 
can be fit to describe the $P_T$- and centrality dependence of RHIC data 
by increasing $\cslpm$ to $0.085$ GeV and $\casw$ to $2.8$ GeV. 
Fig.\ \ref{fig:raa2} shows the results for $R_{AA}$ for 
a central, a mid-central and a peripheral bin using ASW energy loss. 
For each bin three curves are compared to data from PHENIX:
calculations with (i) the average and (ii) event-by-event fireballs using
the old fit values for $\casw$, and (iii) the event-by-event results using 
the newly adjusted parameter $\casw$. sLPM energy loss leads to a 
similar picture. At low transverse momentum the new fits can be matched
perfectly to the original curves from
smooth fireballs, while at high $P_T$ differences can occur, however well
within experimental error bars.
We conclude that the use of smooth fireballs could underestimate the
extracted energy loss coefficient by as much as $50\%$ in the ASW model 
compared to an event-by-event analysis, and still by as much as 25\% in the
sLPM model.
Suppose we do not trust GLISSANDO to capture spatial details of the initial 
collision correctly. We can still make the following model independent
statement: There is an (additional) uncertainty of up to a factor 2 on 
extracted values of $\hat q$ coming from the unknown event-by-event geometry 
of the fireball.

\begin{figure}[tb]
\centerline{\includegraphics[width=\columnwidth]{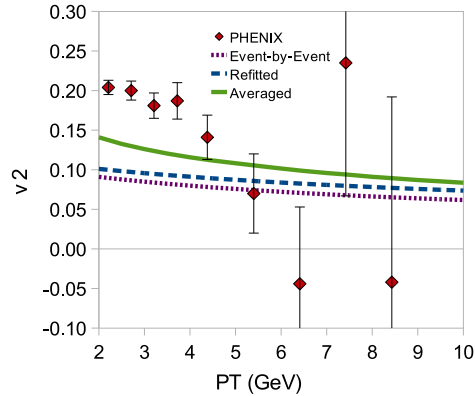}}
\caption{The azimuthal asymmetry $v_2$ of neutral pions as a function of 
$P_T$ for impact parameters around $b=11$ fm compared with data from PHENIX
 \cite{:2009iv}. We show computations in the ASW model using the average event 
(solid line), an event-by-event calculation (dotted line) with 
$\casw$ fitted to the $R_{AA}$ using the average event. We also show
the event-by-event case for $\casw=2.8$ GeV which fits the $R_{AA}$ in the
event-by-event case (dashed line).}
\label{fig:v2}
\end{figure}

Let us proceed to discuss the azimuthal asymmetry $v_2$. As expected from
our analytic arguments the value of $v_2$ decreases for all centrality
bins and for both sLPM and ASW energy loss if event-by-event 
computations are compared to the average fireball with the 
quenching strengths $\cslpm$ and $\casw$ fixed. However, we observe that
readjusting the strength to fit $R_{AA}$ for all centrality bins does
not bring $v_2$ to the level observed for smooth fireballs. This residual
effect increases with impact parameter $b$. Fig.\ \ref{fig:v2} shows
the calculated values of $v_2$ for ASW energy loss for impact parameters
around $b=11$ fm compared to PHENIX data \cite{:2009iv}. At a transverse momentum
of 4 GeV/$c$ the residual suppression of $v_2$ is about 25\%.
This rather deepens the puzzle of $v_2$ calculations which are
routinely underpredicting the experimentally observed values 
at large $P_T$ \cite{Shuryak:2001me}.
On the other hand, an interesting possibility takes shape. 
Looking at $R_{AA}$ alone did not give us any handle on the geometry
of the fireball since a simple rescaling of the energy loss parameters
could absorb the effect. Looking at $v_2$ in addition could in principle
put experimental limits on inhomogeneities in the fireball.

\begin{figure}[tb]
\centerline{\includegraphics[width=\columnwidth]{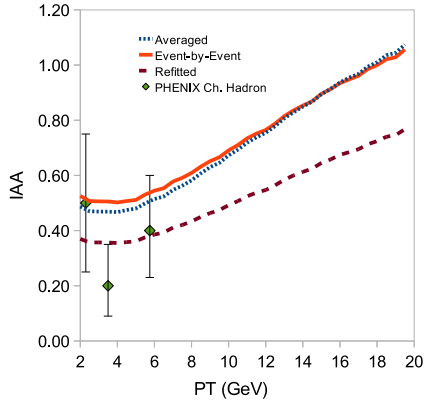}}
\caption{The two-hadron correlation $I_{AA}$ of neutral pion triggers between
  7 and 9 GeV/$c$ and associated charged pions as a function of associated
  $P_T$ for impact parameters around $b=3.2$ fm. Trigger particles are counted 
  in a window from 7 to 9 GeV/$c$. We show computations in the 
  ASW model using the average event (dotted line) with an event-by-event 
  calculation (solid line) with $\casw$ fitted to the $R_{AA}$ using the 
  average event. We also show the event-by-event case for $\casw=2.8$ GeV 
  which fits the $R_{AA}$ in the event-by-event case (dashed line).
  PHENIX data for $\pi^0$-charged hadron correlations and the same trigger
  window are taken from Ref. \cite{Adare:2010ry}.}
\label{fig:jaa}
\end{figure}

Fig.\ \ref{fig:jaa} shows our results for the triggered di-hadron correlation
function $I_{AA}$ in the ASW model for impact parameters around $b=3.2$ fm.
We see that quenching in the average event is larger than for the
event-by-event scenario, analogous to the single hadron case. $J_{AA}$
rises by up to 25\% in the event-by-event case with fixed quenching strength,
but this is almost canceled by the corresponding rise in $R_{AA}$ such that
the modified per trigger yield $I_{AA}$ is almost unchanged.
However, when we use the quenching parameter that fits $R_{AA}$ for 
event-by-event computations to data we observe that the refitting of $R_{AA}$ 
overcompensates the effect for di-hadron quenching in a dramatic fashion. 
$I_{AA}$ from
event-by-event computations is now up to 25\% smaller than for the averaged 
event. This overcompensation could serve as another signature for
inhomogeneities. It is observed for both sLPM and ASW energy loss models.
We conclude that a blend of single and dihadron measurements supplemented 
with $v_2$ measurements can in principle discriminate between different 
scenarios for the density correlation function $R$.
At this point the uncertainties in $I_{AA}$ data are still somewhat large
and quantitative estimates are not yet conclusive.

\section{Summary}

We have shown that realistic fluctuations and inhomogeneities in the
fireball can have significant effects on jet quenching. We tie
the deviation of single particle suppression from that in an average 
fireball to a path integral over the correlation function 
$\langle n(\bfr_1) \rho(\bfr_2)\rangle$ between the fluctuations in
the density of hard processes and the density of the medium.
We predict that for a fixed quenching strength $\hat q(\rho)$ 
$v_2$ should be diminished for a wide class of energy loss models, while 
the sign of the correction to $R_{AA}$ is less obvious and depends on 
details of the correlation function and the energy loss model used.
We expect less suppression for event-by-event jet quenching in the 
limit of very strong, surface-dominated quenching.

We have verified numerically with two energy loss models that at
RHIC energies single hadron suppression $R_{AA}$ is decreased for realistic
event-by-event quenching. On the other hand $v_2$ is decreased as expected.
The quenching strength $\hat q$ as a function of the medium density
$\rho$ can be increased to describe the observed single particle suppression
in event-by-event calculations.
In fact, we can not distinguish, at low transverse momentum, 
between smooth and inhomogeneous fireballs
using the $P_T$- and centrality dependence of $R_{AA}$ alone if the quenching
strength $\hat q(\rho)$ is an adjustable parameter. The quenching strength
has to be increased by up to 100\% which can be interpreted as an
additional uncertainty in the extraction of $\hat q$ from data.

We observe that $v_2$ is still suppressed by up to 25\%, and $I_{AA}$ is
decreased by the same amount even after adjusting the quenching strength to 
fit the data on single hadron suppression. This residual signal of
inhomogeneities can in principle be used for a true tomography which
can measure the degree of initial fragmentation in the fireball.
Of course this is only viable with di-hadron data that has significantly 
smaller error bars, and once theoretical uncertainties from other sources 
in energy loss calculations are under control.

%\ack            
This work was supported by CAREER Award PHY-0847538 from the U.\ S.\ National
Science Foundation, RIKEN/BNL and DOE grant DE-AC02-98CH10886.
E.\ R.\ thanks the Cyclotron Institute at Texas A\&M for its hospitality
and the National Science Foundation for support of the REU program
under award PHY-0647670.

\end{document}